\documentclass[journal=jacsat,manuscript=article]{achemso}

\usepackage{hyperref}
\usepackage{graphicx}% Include figure files
\usepackage{bm}% bold math
\usepackage{color}
\usepackage{soul}
\usepackage{notes2bib}
\usepackage{verbatim}
\usepackage{latexsym}
\usepackage{setspace}
\usepackage[utf8]{inputenc}
\usepackage[english]{babel}
\usepackage{amsmath}
\usepackage{amssymb}
\usepackage{xcolor}
\usepackage{mathtools}

\newcommand{\nc}{\newcommand}
\nc{\nn}{\nonumber}

\newcommand{\onlinecite}[1]{Ref.~[\hspace{-1 ex} \nocite{#1}\citenum{#1}]}

\def\E{\mathcal{E}}

\newcommand{\vast}{\bBigg@{3}}
\newcommand{\Vast}{\bBigg@{5}}
\newcommand{\XYZ}[1]{\color{black} #1}

\usepackage{xr}
\makeatletter
\newcommand*{\addFileDependency}[1]{% argument=file name and extension
  \typeout{(#1)}
  \@addtofilelist{#1}
  \IfFileExists{#1}{}{\typeout{No file #1.}}
}
\makeatother

\author{Yonatan Dubi}
\affiliation[Chem-BGU]{Department of Chemistry, Ben-Gurion University of the Negev, Be'er Sheva, Israel 8410501}
\email{jdubi@bgu.ac.il}

\author{Ieng-Wai Un}
\affiliation[SECE-BGU]
{School of Electrical and Computer Engineering, Ben-Gurion University of the Negev, Be'er Sheva, Israel 8410501}

\author{Yonatan Sivan}
\affiliation[SECE-BGU]
{School of Electrical and Computer Engineering, Ben-Gurion University of the Negev, Be'er Sheva, Israel 8410501}
\title{Distinguishing thermal from non-thermal (``hot'') carriers in illuminated molecular junctions} % coupled to plasmonic electrodes

\begin{document}

\begin{abstract}
The search for the signature of non-thermal (so-called ``hot'') electrons in illuminated plasmonic nanostructures requires detailed understanding of the non-equilibrium electron distribution under illumination, as well as a careful design of the experimental system employed to distinguish non-thermal electrons from thermal ones. Here, we provide a theory for using plasmonic molecular junctions to achieve this goal. We show how non-thermal electrons can be measured directly and separately from the unavoidable thermal response, and discuss the relevance of our theory to recent experiments.
\end{abstract}

\section{Introduction}
When a plasmonic nanostructure is continuously illuminated, two things happen simultaneously. First, the system unavoidably heats up, and second, non-thermal (so-called ``hot'') carriers (NTCs) are being generated. The latter effect leads to an electron (and hole) distribution that deviates from the equilibrium (Fermi-Dirac) distribution~\cite{Seidman-Nitzan-non-thermal-population-model,Dubi-Sivan}. There has been growing interest in understanding the interplay between these two processes; non-thermal carriers were suggested to be beneficial for various applications, including, notably, photodetection~\cite{Uriel_Schottky,Uriel_Schottky2,Moskovits_hot_es,Valentine_hot_e_review,Giulia_Nat_Comm_2018} and plasmon-driven chemistry~\cite{plasmonic-chemistry-Baffou,Baffou_solvothermal,plasmonic_photocatalysis_Clavero,wei2018quantum,Jain_viewpoint,Baffou-Quidant-Baldi,Dubi-Sivan-APL-Perspective}. However, the latter application is very sensitive to the temperature of the system. Thus, distinguishing between NTCs and thermal effects is crucial for understanding how (and if at all) NTCs can be used to, e.g., catalyse certain reactions. 

Various indirect ways were proposed to determine the non-thermal electron population. These include primarily the study of the ultrafast dynamics of the metal permittivity via transient reflectivity measurements~\cite{non_eq_model_Lagendijk,delFatti_nonequilib_2000,nt_electrons,Langbein_PRB_2012,Stoll_review} and of the photoemission~\cite{Bauer,Aeschliman_e_photoemission_review,Lienau_e_photoemission_tip,Petek-PRL-2019} following illumination by an ultrashort pulse. Under such conditions, a relatively large number of high energy non-thermal electrons are being generated, but it is not clear when and how to separate those from mere heating. In contrast, under continuous wave illumination, it is clear that the information about heating is characterized by Fermi-like distribution of carriers close to the Fermi level, whereas the NTCs reside further away from the Fermi level in nearly-flat ``shoulders'', see~\onlinecite{Dubi-Sivan,Dubi-Sivan-Faraday}. However, the practical separation between thermal and non-thermal carriers is very challenging, because the number of the high excess-energy non-thermal electrons is many orders of magnitude smaller compared to the number of thermal (i.e., low excess energy) carriers. Thus, the various attempts made to directly quantify the contributions of these two types of charge carriers (e.g., to chemical reaction rates in the context of plasmon-assisted photocatalysis), frequently fail~\cite{anti-Halas-Science-paper,Y2-eppur-si-riscalda,anti-Halas-NatCat-paper,R2R}, because the control thermocatalysis (light off) experiments must reproduce exactly the temperature profile of the photocatalysis (light on) experiments~\cite{Y2-eppur-si-riscalda,Dubi-Sivan-APL-Perspective}, a task which is nearly impossible (although progress has been made in this direction~\cite{Baldi-ACS-Nano-2018,yu2019plasmonic,Liu-Everitt-Nano-research-2019,Giulia_2019,Boltasseva_LPR_2020,Langhammer-acs-nano-2021}). 

In an attempt to circumvent this problem, Reddy, Wang and co-authors~\cite{reddy2020determining} (referred to as RW20 hereafter) recently suggested measuring directly the NTC distribution by coupling a plasmonic Au film to a molecular junction (schematically described in Fig.~\ref{fig:Delta_J1}(a), adapted from~RW20). By measuring the I-V curves through the molecular junction (MJ) under illumination and in the dark, these authors assess directly the effect of illumination on the electronic distribution in the illuminated Au electrode. 

% % PL - Ono, Lupton - ultrafast and not really aimed at studying the e population. Sheldon - aims at the population but misinterprets his results... so nothing on PL here.

The theoretical foundations to describe the non-thermal electron distribution as well as the current through such a plasmonic MJ were laid in~\onlinecite{Seidman-Nitzan-non-thermal-population-model}. We recently extended the first part of the model of~\onlinecite{Seidman-Nitzan-non-thermal-population-model} to provide a complete model of the electron non-equilibrium under CW illumination~\cite{Dubi-Sivan,Dubi-Sivan-Faraday}. Here, we extend the second part of the theory of~\onlinecite{Seidman-Nitzan-non-thermal-population-model}, namely, transport through a MJ coupled to two electrodes, by combining the standard Landauer theory of transport through MJs~\cite{cuevas2010molecular,diventra2008electrical} and the analytic form for the electron non-equilibrium distribution of an illuminated metal~\cite{Dubi-Sivan-Faraday}, and suggest a scheme that can be used to measure the non-thermal electron distribution in the presence of strong heating and even large thermal gradients (i.e., regardless of the temperature distribution). Using this formulation we shed new light on the experimental results of Reddy, Wang {\em et al.},~\cite{reddy2020determining} demonstrating that it is possible that they indeed were able to distinguish (probably for the first time) non-thermal electrons from thermal ones, but not in the way interpreted in the original manuscript. %\XYZ{Specifically, we show that the signature of non-thermal electrons is absent from the central result of RW20 (which can be explained via a purely thermal model), yet, it is reflected in a additional experiments they performed. }
{\XYZ Specifically, we show that the signature of non-thermal electrons is weak in some of the key measurements of RW20, which can be explained alternatively via a purely thermal model. However, we point out that the signature of non-thermal electrons is dominant in some other experiments that have been performed by the authors of RW20.}

\section{Results and Discussion}

\subsection{General Theory}
Current through a molecular junction is typically described by the Landauer formula~\cite{cuevas2010molecular,diventra2008electrical}, which relates the total current to the electronic transmission function $\mathcal{T}(\E)$ and the electrodes' electron distribution functions, $J = \frac{2e}{h} \int d\E \mathcal{T}(\E) \left(f_{top} - f_{bottom}\right)$; here, $e$ is the electron charge, $h$ is Planck's constant, and $f_{top/bottom}$ represent the electron distribution of the top/bottom electrodes (representing the STM tip and the  Au slab, respectively); see Fig.~\ref{fig:Delta_J1}(a). At equilibrium, $f_{top} = f_{bottom} = f^T(\E,T_e)$, where $f^T(\E,T_e) = \left(1 + \exp\left(\frac{\E - \mu}{k_B T_e}\right) \right)^{-1}$ is the thermal (Fermi-Dirac) distribution, $T_e$ the (electron) temperature, $\mu$ the chemical potential and $k_B$ the Boltzmann constant. The transmission function is typically described by a Lorentzian, $\mathcal{T}(\E) = \frac{\Gamma^2}{\Gamma^2 + (\E - \E_0)^2}$, where $\E_0$ is the energy of the frontier molecular orbital and $\Gamma$ is the level broadening (this is the so-called wideband approximation~\cite{cuevas2010molecular,diventra2008electrical}). Notably, the Landauer formula implicitly assumes that the electron and temperature distributions on the electrodes is uniform. We point that this approach ignores the direct effect of illumination on the electrons in the molecule, which is justified because the illumination is off-resonance with the molecular HOMO-LUMO gap (and also with $\E_0 - \mu$ which may affect transport~\cite{zhou2018photoconductance}), and hence the effect is negligible. It also disregards electron interactions in the molecule, which may shift the resonance position, but steal lead to Lorentzian-like transmission \cite{Evers2021,thoss2018perspective}. These approximations were also used in, e.g., RW20. % (or can reasonably-well uniformized)%. Alternatively, it relies on the local values of the distribution and temperature {\bf Yoni - approve/modify please}. 

With equal distributions on the two electrodes, the current vanishes. As bias voltage $V$ is applied, the chemical potentials shift such that $\mu_{top}-\mu_{bottom} = eV$ and current flows through the junction(we hereafter treat voltage in units of electron-volt). Current can also be driven by a temperature difference (i.e., setting $T_{e,top} \neq T_{e,bottom}$), generating a thermo-electric effect~\cite{cuevas2010molecular,diventra2008electrical,dubi2011colloquium}. 

For the current to give an indication on the non-thermal electron distribution, it needs to be generated using optical illumination in an asymmetric manner, such that electric field felt by the electrodes is substantially different, and as a result, $f_{top} \neq f_{bottom}$~\cite{vadai2013plasmon,arielly2011accurate,wang2016molecular,banerjee2010plasmon}. The MJ geometry enables this naturally; indeed, electromagnetic and thermal simulations of a generic MJ geometry reveals significant differences in the local fields and temperature across the MJ, see SI section 3.1.  To evaluate this effect, we follow~\onlinecite{Dubi-Sivan-Faraday} (specifically its Appendix A), where it was shown that under continuous illumination (i.e., by monochromatic light at frequency $\omega_L$), the electron distribution in a Drude metal is (to an excellent approximation)\footnote{We point that while other formulations of the electron non-equilibrium exist, their limitations (see discussion in~\onlinecite{Dubi-Sivan,Dubi-Sivan-Faraday}) make them unsuitable for separating thermal from non-thermal effects.}
\begin{equation} \label{Eq:f_light}
f(\E;T_e,\omega_L,|E|^2) = f^T(\E;T_e) + \delta_E(\E) \left[f^T(\E + \hbar \omega_L;T_e) + f^T(\E - \hbar \omega_L;T_e)\right]~~,
\end{equation}
where $\delta_E$ measures the population of non-thermal carriers, or more precisely the deviation of the distribution function from its equilibrium form, and is given by 
\begin{eqnarray}\label{eq:R}
\delta_E(\E,T_e) &\equiv& \left|\frac{E}{E_{sat}}\right|^2, \quad \quad \left|E_{sat}(\E,T_e)\right|^2 \equiv \frac{1}{\tau_{e-e}(\E,T_e) R}. % \\ \nonumber R &\equiv& \frac{4 \epsilon_0 \epsilon_m''(\omega_L)}{3 \hbar n_e} \frac{\E_F}{\hbar \omega_0},
\end{eqnarray}
Here, $|E|^2$ is the local electric field intensity and $R$ is a constant that depends on the (imaginary part of the) metal permittivity at the laser frequency $\epsilon_m''$, and electron density $n_e$ but not on $T_e$~(see SI Section~3.1)\footnote{Note that the expression for $R$ in the original derivation~\cite{Dubi-Sivan-Faraday} had a small typo; it was corrected in an errata, and noted in~\onlinecite{Sivan-Dubi-PL_I}. }; $\tau_{e-e}(\E,T_e) = \{K \left[(\pi k_B T_e)^2 + (\E - \mu)^2 \right]\}^{-1}$ is the $e-e$ collision rate, for which we adopt the Fermi liquid theory, $K$ being the $e-e$ scattering constant~\cite{Quantum-Liquid-Coleman}. The importance of high energy non-thermal (i.e., ``hot'') carriers, via the change in the distribution due to hot carriers, can be simply quantified by evaluating $\delta_E(1) = \delta_E(\E - \mu = 1eV)$, namely, the value of $\delta_E$ at an energy 1eV above the Fermi energy, which captures the deviation from a thermal distribution at high energies. Notably, the solution for the non-equilibrium distribution~\eqref{Eq:f_light} is also obtained under the assumption of uniform (or averaged) field.

Evaluating the current through an asymmetrically illuminated MJ can now be easily done by setting $f_{top} = f(\E + V;T_{e,top},|E_{top}|^2),~f_{bottom} = f(\E ;T_{e,bottom},|E_{bottom}|^2)$ (assuming that the bottom electrode is grounded\cite{reddy2020determining}) and plugging these distributions into the Landauer formula for the current. To isolate the contribution of illumination, we follow the authors of~\onlinecite{reddy2020determining} who subtracted from the current under illumination the current in the dark (both as a function of voltage). In this case, the contribution to the current from illumination is simply 
\begin{equation} \label{J_light}
\Delta J_{light} = J_{light}(V,T_{e,top},T_{e,bottom},|E_{top}|^2,|E_{bottom}|^2) - J_{dark}(V,T_{dark}). 
\end{equation}

Eq.~\eqref{J_light} implies that the current is determined by the electric field and electron temperature rise induced by the illumination at both electrodes. % Note that the electron temperatures $T_{e,top},T_{e,bottom}$ are also by themselves determined by the electric field~\cite{Dubi-Sivan}. 
We use numerical simulations to evaluate these quantities, and then we evaluate $\Delta J_{light}$ for various molecular and illumination conditions. We show below that, depending on the properties of the molecular system, $\Delta J_{light}$ has a dominant feature coming from either the temperature difference or the NTC contribution. %These calculations are supplemented by evaluation of the electric field in the MJ as described above. 
% We show that excellent fits to the data of RW20 can be achieved, corroborated by the electromagnetic calculations. 

%{\bf re-think the context for this paragraph + refer to plots in the SI} %the following unnecessary - Due to the few nm size of the MJ, one would expect that any {\em incident} electric field is nearly uniform in the vicinity of the MJ, thus, affecting the two side of the MJ (i.e., both slab and STM tip). **********

%One would expect the field distribution to strongly depend on the illumination. Specifically, if the incident field has a vertically-oriented electric field component (e.g., as for SPP-based illumination in RW20~\cite{reddy2020determining}), the strongly-localized (dipolar) modes of the nano-cavity created by the tip and slab would be excited, giving rise to a strong field enhancement. If, in addition, the STM tip is sharp, the local field in it would be higher than in the slab. For example, our simulations show a 3-4 times stronger field compared to the slab for a tip of $1$nm radius of sharpness; {\bf ?? and for a wider one??; now if the tip is flat, then, there should be no signal for L2...}. In contrast, if the vertical field component is absent from the illumination (e.g., as for illumination from bottom, as in the control experiments of RW20), then, there is no coupling to the dipolar modes and only the minor polarization is excited. In this case, the field in the slab turns out to be higher. 

\subsection{Results - nearly-resonant molecules}
\begin{figure}
\centering{\includegraphics[width=0.7\textwidth]{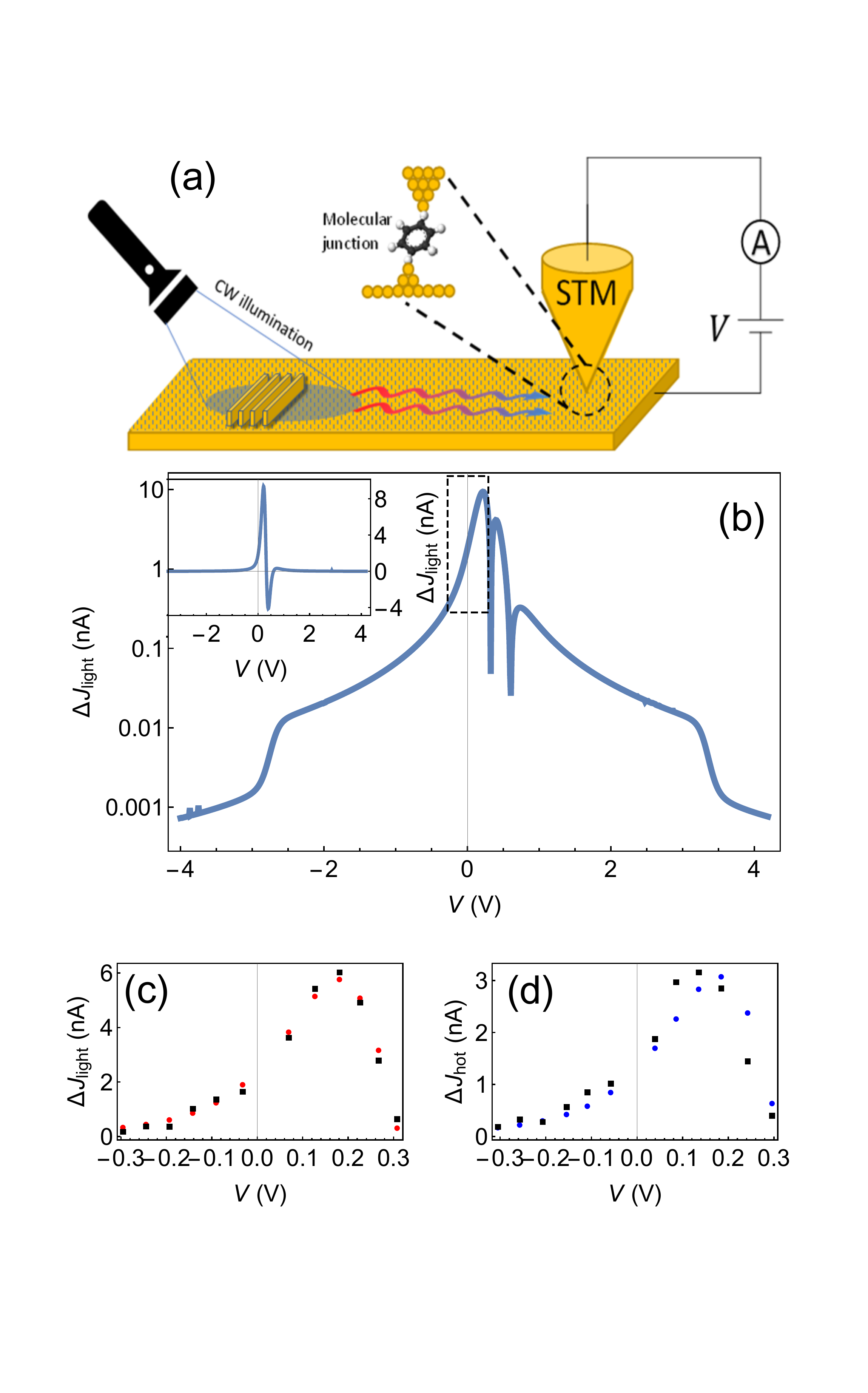}}
\vskip -3truecm \caption{(Color online) (a) Schematic depiction of the experimental setup, comprising an Au slab with nanofabricated gratings, and a molecular junction (MJ) formed between the slab and an STM tip. Surface plasmons are excited in the slab by CW illumination of the grating, and propagate towards the MJ (wiggly arrows). (b) $\Delta J_{light}$~\eqref{J_light} as a function of bias voltage (log scale) for a molecular junction with resonance close to the Fermi level, with parameters $\E_0 = 0.15$ eV and $\Gamma = 10$ meV, and setting $T_{e,bottom} = 305$ K,  $T_{e,top} = T_{dark} = 300$ K and $\delta_E(1) = 10^{-5}$) (see text for further details). Near zero bias, the most dominant feature is the thermal contribution to the current. Spikes in the plot correspond to voltages at which the current changes sign. The dashed line shows $\Delta J_{light}$ without illumination, i.e., all parameters the same except $\delta_E(1) = 0$, thus showing only the thermal response. The dashed black square shows the region of measurement in RW20 (note the match between solid and dashed lines in this region, marking that non-thermal effects are negligible there). Inset: same plot in linear scale. (c)-(d) Fits to experimental data of Reddy {\sl et al.,}~\cite{reddy2020determining} for (c) $6$nm and (d) $13$nm thick slabs, assuming the tip is not heated, but the slab right under it is. Molecular parameters were set to $\E_0 = 0.155$eV and $\Gamma = 0.057$eV. Red and blue circles are the experimental data and black squares are the fits, yielding $\delta T = 2.21$ K and $1.14$ K for (c) and (d) respectively (see text for further discussion on these parameters). } \label{fig:Delta_J1}
\end{figure}

We start with addressing a "nearly-resonant" molecule, i.e., a molecule for which the orbital energy is close to the electrodes' Fermi level (generally, it can represent either the HOMO or LUMO levels in the molecule, depending on the molecular moiety). In Fig.~\ref{fig:Delta_J1}(b), we plot (the log of the absolute value of) $\Delta J_{light}$ as a function of voltage for a molecular junction, taking relevant parameters for a typical MJ, $\E_0 = 0.15$ eV and $\Gamma = 10$ meV; the relatively small value of $\E_0$ makes this molecule "nearly-resonant" (with respect to the electrodes' Fermi level). We take $\delta_E(1) = 1.5 \times 10^{-5}$ (which corresponds to the value we calculate from microscopic consideration, see SI Section 3) and $\hbar \omega_L = 1.48$ eV (corresponding to $833$ nm wavelength); note that the chosen wavelength is sufficiently longer than the $\sim 1.77$eV threshold for interband transitions~\cite{Rosei_Au_diel_function}, thus, validating the use of the analytic solution for Drude metals (Eqs.~(\ref{Eq:f_light})-(\ref{eq:R})). We further set $T_{dark} = 300$ K, and assume that the illuminated slab under the STM tip is heated by $5$ K (i.e., $T_{e,bottom} = 305$ K). These parameters are close to those presented in RW20. We also assume (as in RW20) that the STM tip itself is not heated ($T_{e,top} = T_{dark}$). 

In this case, since the non-thermal part of the distribution is negligible for realistic values of voltage bias ($\sim \pm 0.3 V$ in RW20), the contribution of the light to the current is given by\cite{reddy2020determining}
\begin{eqnarray} \label{J_light_near_res}
\Delta J_{light} &=& \frac{2e}{h} \int d\E \mathcal{T}(\E + V/2)\left\{ f^T(\E ;T_{e,bottom}) - f^T(\E + V;T_{e,top})  \right. \nonumber \\ & & ~~ \left. - \left(f^T(\E;T_{dark}) - f^T(\E + V;T_{dark})\right)\right\}~~\nonumber\\
&\simeq& \frac{2e}{h} \int d\E \mathcal{T}(\E + V/2)\left\{ f^T(\E ;T_{e,bottom}) - f^T(\E;T_{dark}) \right\}. 
\end{eqnarray}

$\Delta J_{light}$ is plotted in Fig.~\ref{fig:Delta_J1}(b), showing two prominent features. The first is that the strongest current occurs at low voltages, close to $\E_0$; this is due to a thermal effect (coming from the heating of the slab under illumination), centered a width $\sim \Gamma$ around the orbital resonance $\E_0$~\cite{dubi2011colloquium,cuevas2010molecular}. Specifically, the absolute value of $T_{dark}$ has only a minor effect on the results of Fig.~\ref{fig:Delta_J1}(b); it is the difference between the temperature of the bottom electrode upon illumination and its temperature in the dark which is responsible for the large changes near $V = 0$. In that regard, this thermal feature is not a ``thermo-electric'' effect, meaning that increasing the tip temperature (and hence, modifying the temperature difference between the electrodes) has a negligible effect on the current difference $\Delta J_{light}$. The second feature is the onset of non-thermal electron ``shoulders''~\cite{Dubi-Sivan} in $\Delta J_{light}$, which extend into the high-V regime, where the thermal effect becomes small. % The relative importance of thermal and non-thermal effects can readily be seen in the inset to Fig.~\ref{fig:Delta_J1}(b), which shows the same data on a linear scale. Fig.~\ref{fig:Delta_J1}(b) also shows "spikes" or dips in $\Delta J_{light}$. These are simply changes of sign in $\Delta J_{light}$ which appear as spikes in the log-plot, and do not have any special physical meaning.  
{\XYZ Fig.~\ref{fig:Delta_J1}(b) shows "spikes" or dips in $\Delta J_{light}$. These are simply changes of sign in $\Delta J_{light}$ which appear as spikes in the log-plot, and do not have any special physical meaning (see the inset to Fig.~\ref{fig:Delta_J1}(b), which shows the same data on a linear scale). }

% {\bf don't you want to say anything about the "weak" coupling of the molecule which kills the thermo-electric effect? that we also think that there is no thermo-electric effect but because of different reasoning?} 

Armed with these insights from Fig.~\ref{fig:Delta_J1}(b) (mainly that at low bias voltages thermal effects are dominant for a nearly-resonant molecule), we used Eq.~(\ref{J_light}) to fit the measured values of $\Delta J_{light}$ of RW20 for their L1 molecule (a charge-transfer complex of quaterthiophene (T4) and tetracyanoethylene (TCNE) with terminal thiophenes containing gold-binding methyl sulfides; data extracted from the supplementary material (Fig.~S16A in RW20)). We use data from two sets of measurements in RW20, using Au slabs with thicknesses $6$ and $13$nm. We set as free parameters both the molecular parameters $\E_0$ and $\Gamma$ and the local temperature in the Au slab segment directly under the STM tip, $T_{e,bottom} = T_{dark} + \delta T$, where $T_{dark}$ is the ambient temperature and $\delta T$ is the excess temperature of the slab above ambient; we again assume that the top electrode remains unheated. We simultaneously fit both data sets with the same $\E_0$ and $\Gamma$, leaving only $\delta T$ to vary between experiments. 

In Fig.~\ref{fig:Delta_J1}(c)-(d) we plot the experimental data and a best fit to Eq.~(\ref{J_light}) (black squares) with common parameters $\E_0 = 0.155$eV and $\Gamma = 0.057$eV, with $\delta T = 2.21$K and $1.14$K for the $6$ and $13$nm slabs (red and blue circles, respectively). These fits indicate that a thermal origin for the experimental results is plausible. Importantly, associating $\Delta J_{light}$ with a thermal effect naturally explains the two energy scales appearing in the data (the position of the peak $\E_0$ and the linewidth $\Gamma$), which, in contrast, cannot be associated with any feature of NTCs (e.g., in Eq.~(\ref{eq:R})). Note that while these same results were interpreted in RW20 as the ``hot'' carrier contribution, no direct fit between theory and experiment was presented in RW20. In fact, while it is customary to refer to ``hot'' electrons in the energy range close to the Fermi energy, the match of our Drude-based analysis to the measured data confirms that in this energy range, there is predominantly only thermal electrons, i.e., electrons that are characterized by a Fermi-Dirac distribution (with a temperature potentially higher than ambient) whereas the number of true non-thermal electrons is much smaller (proportional to $\delta_E(\mathcal{E} \sim \mathcal{E}_0 \pm \Gamma) \ll 1$)\footnote{An exception to this statement may occur in the presence of interband transitions, and due to the minute deviation of the distribution from equilibrium discussed in Fig. 2(a) of~\onlinecite{Dubi-Sivan}. }. {\XYZ However, we also note that the signature of true non-thermal electrons would have been visible for higher voltages, via the pronounced lower current peak at energies above $\E_0$, compare the solid and dashed lines in Fig.~\ref{fig:Delta_J1}(b); unfortunately, this regime was not accessible experimentally.}

Finally, we note that our best fit parameters mentioned above yield a rather small temperature rise in the Au slab below the STM tip ($\sim 1-2$K). In SI Section 3.2, we explain why these values, as well as the negligible heating assumed in both RW20 and our analysis, are likely to match numerical thermal simulations of the experimental system for the (hard-to-simulate) several micron-wide STM tip, as indeed were supposedly used in RW20. Moreover, we show in SI section 2 that the data of RW20 can be fitted even to different slab temperatures. 

Despite the compelling match of the fit and our simulations and analysis, a few words of caution are in order. Our fits seem to be somewhat inconsistent with some of the data presented in RW20. For instance, the fitted values for $\E_0$ and $\Gamma$ are different from those obtained in RW20 (although the estimates in RW20 are also inconsistent with some of their data, see SI Section 1), but are consistent with earlier measurements of a similar junction~\cite{wang2019charge}. Our results also show a discrepancy with the control experiments performed in RW20, where current in the dark was measured at increased temperatures (RW20, Fig.~S12), and showed essentially no sign of being temperature sensitive. On the other hand, these control experiments are inconsistent with the formalism and parameters used within RW20. Indeed, plugging the RW20 parameters into a calculation of the current reveals a strong dependence on temperature (for the same parameters of RW20 Fig.~S12), yet no such temperature dependence was measured; see SI Section 3.3 for further discussion.

\subsection{Off-resonance molecule}
The difficulty in distinguishing the thermo-electric and non-thermal electron contributions to $\Delta J_{light}$ stems from the two following points: (i) one cannot measure directly the sample temperature under the STM tip, and (ii) because of the relatively small $\E_0$ (the nearly-resonant nature of the molecule), and the fact that thermal current response is limited to a width $\sim \Gamma$ around the molecular orbital $\E_0$~\cite{dubi2011colloquium,cuevas2010molecular}, the thermal effect is most important for low (hence realistic) voltages, i.e., for molecular resonances around the Fermi level. 

Overcoming the first point is very challenging. However, overcoming the second point is actually quite easy. In a molecular junction where the molecular orbital energy $\E_0$ is far from the Fermi level but in resonance with the illumination energy (i.e., $\E_0 \sim \hbar \omega_L$), low voltage measurements of $\Delta J_{light}$ will {\sl only} show the non-thermal part of the distribution. This is because, as is evident from Eq.~(\ref{Eq:f_light}) and the Landauer formula, the non-thermal features extend to a distance $\hbar \omega_L$ from the molecular resonance, and will thus be prominent at low voltages.% {\bf we have a prediction of the width of the thermal range - it shows that the non-thermal parts starts at 0.23eV...

In this case, since the thermal contribution is negligible, the contribution of the illumination to the current is given by
\begin{equation} \label{J_light_off}
\Delta J_{light}\approx \frac{e}{h} \int d\E \mathcal{T}(\E + V/2)\left\{\delta_{E,bottom} f^T(\E - \hbar \omega_L,T_{e,bottom}) -\delta_{E,top} f^T(\E + V - \hbar \omega_L;T_{e,top})\right\}~~, 
\end{equation}
where $\delta_{E,bottom/top}$ are derived (using Eq.~(\ref{eq:R})) from the local field on the bottom/top electrodes. Note that in Eq.~(\ref{J_light_off}) only electrons are considered, i.e., the correction to the distribution corresponding to holes ($\sim f(\E + \hbar\omega_L)$) is disregarded, since holes will shift the distributions further away from the resonance and will not contribute to $\Delta J_{light}$.

\begin{figure}[t]\label{fig-deltaJ2}
\centering{\includegraphics[width=0.8\textwidth]{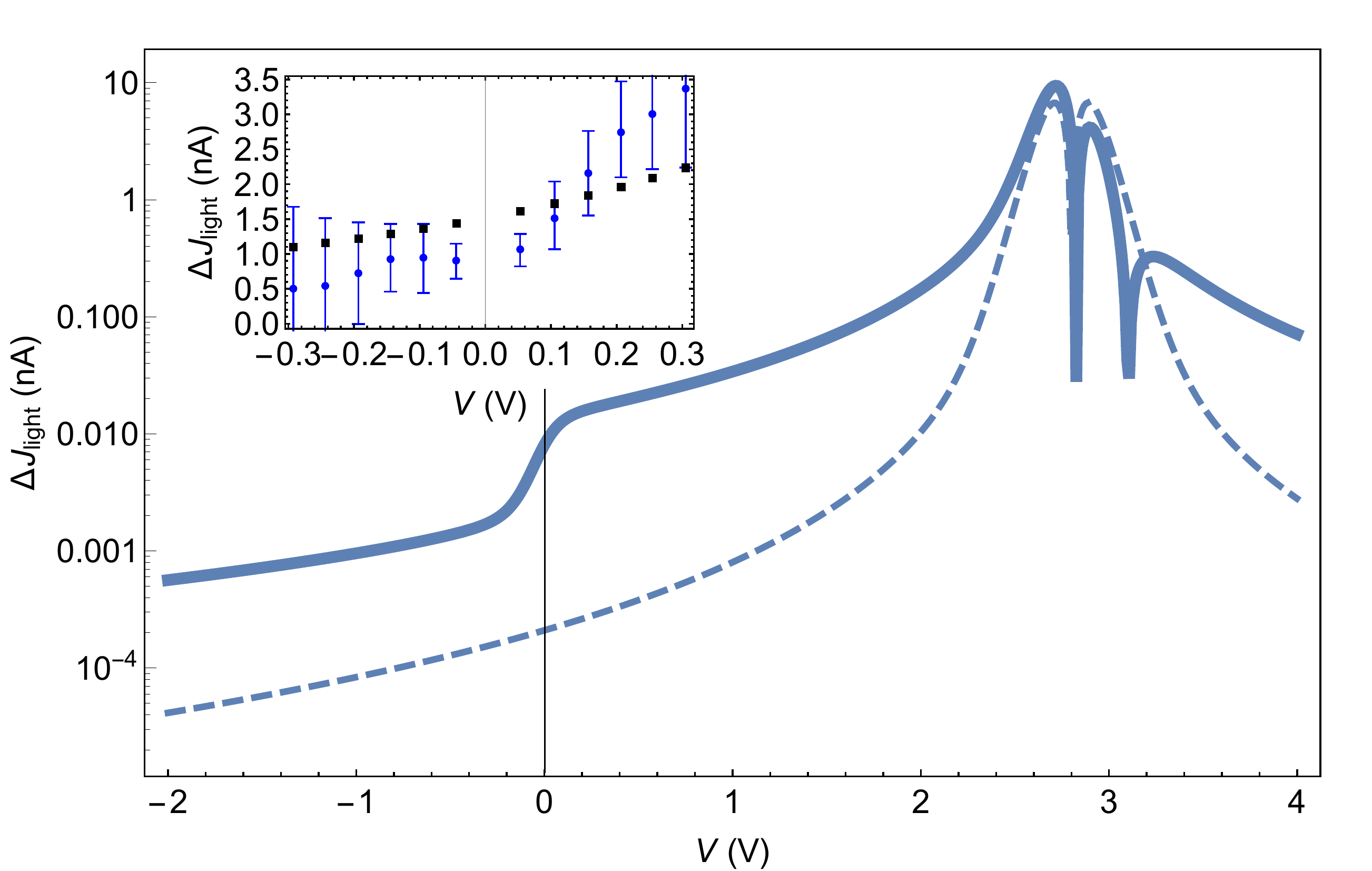}}
\caption{(Color online) $\Delta J_{light}$ (see Eq.~\eqref{J_light}) as a function of bias voltage (log scale) for a molecular junction with molecular resonance $\E_0 = 1.4$ eV far to the Fermi level, but in resonance with the illumination frequency. We take $\delta_{E,top}(1) = 1 \times10^{-5}, \delta_{E,bottom}(1) = 0$, and all other parameters are the same as for L1 (see text for details). Dashed line: same, without any NTCs, showing that the thermal contribution in these case is negligible near the Fermi level. Inset: Fit between experimental data for molecule L2 and a $6$nm slab (Fig. S18 of RW20, blue circles) and theory (black squares), demonstrating that this measurement may indeed be an indication for NTCs. The molecular parameters $\Gamma = 0.18$eV and $\E_0 = 0.77$eV and $\delta_{E,top}(1) = 8 \times 10^{-6} $ (and keeping $\delta_{E,bottom}(1) \approx 0.1 \delta_{E,top}(1)$) (see text for details). 
} \label{fig:Delta_J2}
\end{figure}

In Fig.~\ref{fig:Delta_J2} we plot (the log of) $\Delta J_{light}$ as a function of bias voltage (as in Fig.~\ref{fig:Delta_J1}(b)) for the case of $\E_0 = 1.4$ eV, at resonance with $\hbar \omega_L$ (all other parameters are the same as for L1, solid line), for $\delta_{E,top}(1) = 1\times10^{-5}, \delta_{E,bottom}(1) = 0$. It is clear that now the thermal feature only appears at high (hence, inaccessible) voltages, while at low voltages, the non-thermal electron ``shoulder'' provides the prominent contribution to the current. For comparison, the dashed line shows $\Delta J_{light}$ for $\delta_{E,bottom}(1) = 0$ (i.e., no non-thermal electrons), demonstrating the orders-of-magnitude larger contribution of non-thermal electrons at low voltages. Importantly, there are also qualitative differences with respect to a nearly-resonant molecule like L1, most prominent is the fact that $\Delta J_{light}$ does not change sign, a feature which can be easily recognized experimentally (indeed, see below). 

A somewhat similar experiment was, in fact, conducted in RW20, using a MJ with a 1,4-benzenediisonitrile molecule (dubbed L2 in RW20). This molecule has a LUMO level which is far from the Fermi level, $\E_{LUMO} - \mu \sim 0.77$ eV~\cite{lee2013heat}, and thus is somewhat similar to the situation described above. In the SI to RW20 (Fig.~S18), the authors plot $\Delta J_{light}$ vs. the bias voltage. We use these data to fit Eq.~(\ref{J_light}), and find that for this molecular energy indeed the thermal contribution is negligible for that range of voltages, and that the data can be fitted very well (within the experimental error) with the contribution coming solely from the non-thermal part of the distribution. This is shown in the inset to Fig.~\ref{fig:Delta_J2} where the experimental data (blue points) and the theoretical points (black squares) are shown. The molecular parameters $\Gamma = 0.18$eV and $\E_0 = 0.77$eV are taken from~\onlinecite{lee2013heat}, and the only fit parameters are $\delta_{E,bottom}(1)$ and $\delta_{E,top}(1)$.

To reduce the number of fit parameters further, we have conducted numerical simulations of the electric field under the experimental conditions (see SI Section 3.1). These simulations show that the field in the tip is actually {\sl larger} than the field in the slab by a factor of $\sim 3.2$, due to the plasmonic enhancement around the tip. 

In the inset to Fig.~\ref{fig:Delta_J2} we plot the experimental data of RW20 (blue points, along with the experimental error bars), and the theoretical $\Delta J_{light}$ as a function of bias voltage, where $\delta_{E,top}(1) = 8 \times 10^{-6} $ (and keeping $\delta_{E,bottom}(1) \approx 0.1 \delta_{E,top}(1)$) is found to provide the best fit value. This fit between data and theory provides further experimental corroboration to Eq.~(\ref{J_light}), providing what is, to the best of our knowledge, the first direct measurement of the steady-state non-thermal contribution to the electron distribution. These values for $\delta_E$ match the calculated values for the electric field (see SI Section 3.1) and are in good correlation with reported ``hot'' electron efficiencies in plasmon-assisted photocatalysis experiments~\cite{Baldi-ACS-Nano-2018,Jain_Nat_Chem_2018,Giulia_2019}.

However, somewhat unintuitively, the experimental data of the L2 molecule in RW20 can also be explained by simple heating. Indeed, by setting the $\delta_E$'s to be zero and assuming $\delta T_{e,bottom} \sim 20$K we obtain a fit essentially similar to that shown in the inset of Fig.~\ref{fig:Delta_J2}; note that this is not a contradiction with Fig.~\ref{fig:Delta_J2}, because the energy of molecule L2 obtained from the fit is much lower (i.e., not as far from resonance) compared with the value used for the illustration of Fig.~\ref{fig:Delta_J2}; in that respect, the experiment in RW20 involves an intermediate case, whereby the molecular energy is only partially off-resonance. One is thus facing a situation where both NTCs and thermal effects can reproduce the experimental data.

\begin{figure}[h]
\centering{\includegraphics[width=0.8\textwidth]{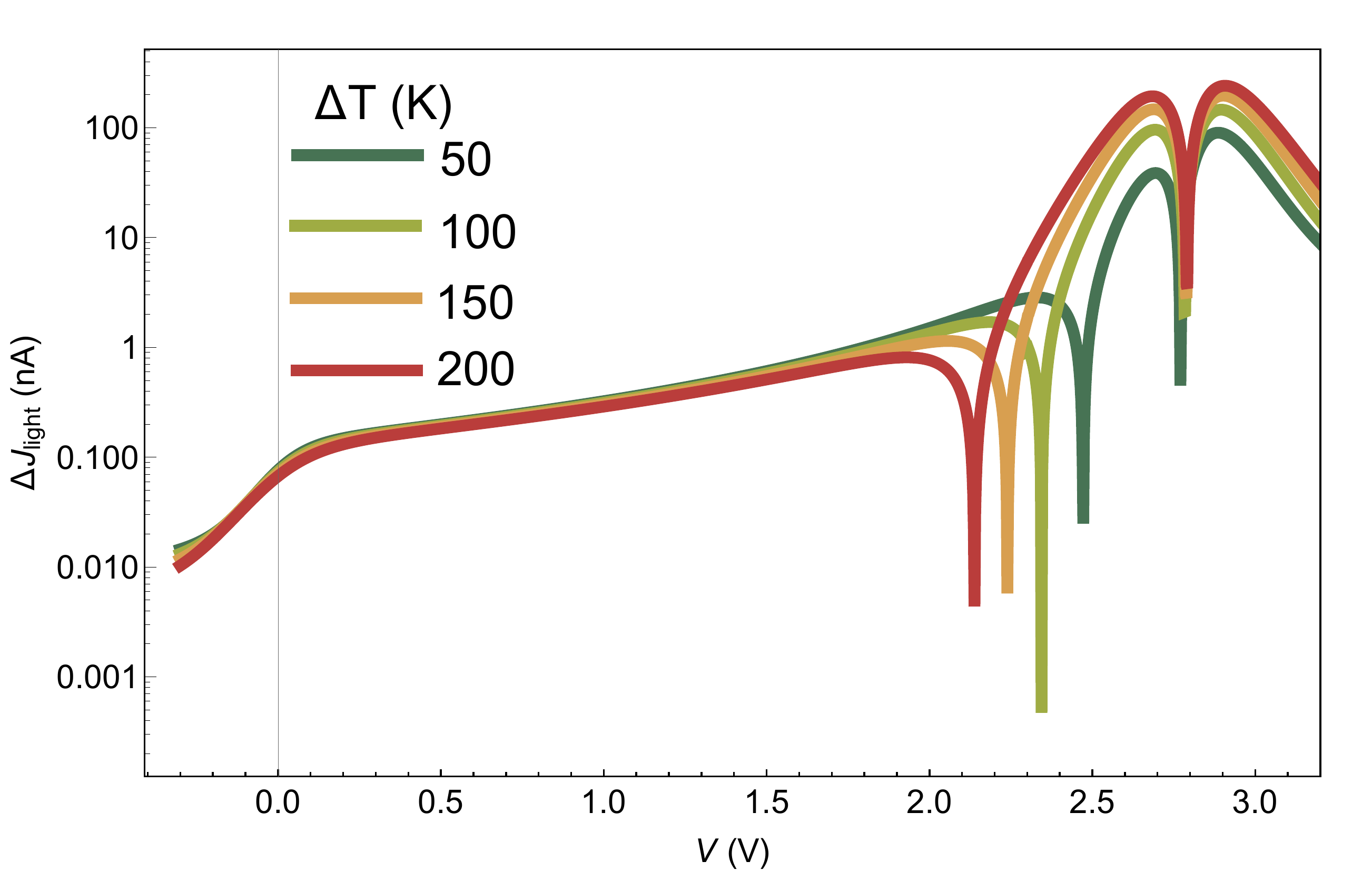}}
\caption{(Color online) $\Delta J_{light}$~\eqref{J_light} as a function of bias voltage (log scale) for different values of electron temperature, $T_{e,bottom} = T_{dark} + 50, 100, 150, 200$ K (see text for other parameters).The thermal effect (which appears close to the molecular resonance strongly depends on temperature, while the NTCs contribution is temperature independent, thus providing a simple way for discriminating between the two effects. } \label{fig:Delta_J_dT}
\end{figure}

Fortunately, unlike the situation in plasmon-assisted photocatalysis experiments (e.g.,~\onlinecite{anti-Halas-Science-paper}), the MJ setup offers a simple way to further discriminate between the two effects, by simply measuring $\Delta J_{light}$ (for the off-resonance molecule) as a function of the ambient temperature. The reason is that while the thermal contribution would be strongly influenced by the temperature, the NTCs would not (see Eq.~(\ref{Eq:f_light}) and~\onlinecite{Dubi-Sivan-Faraday}. In order to demonstrate this, in Fig.~\ref{fig:Delta_J_dT}, we plot $\Delta J_{light}$ as a function of bias voltage for $T_{e,bottom} = T_{dark} + 50, 100, 150, 200$ K, and as can be seen, the low-voltage NTC contribution to $\Delta J_{light}$ is essentially unchanged. Another way to overcome the situation is to choose a molecular junction which has frontier orbitals which are even further away from the Fermi level of the electrodes, such as Au-benzenedithiol-Au junctions, where the HOMO level is $\sim 1.2$eV away form the Fermi level~\cite{reddy2007thermoelectricity}. Under such circumstances, the thermal effect is expected to contribute only a tiny fraction of the NTC contribution to $\Delta J_{light}$. %{\bf Unfortunately, this approach is difficult, because the current will drop rapidly as the molecule resonance is moved away from the Fermi energy (compare e.g., Fig.~\ref{fig:Delta_J2} with its inset).} %In the inset of Fig.~\ref{fig:Delta_J_dT} we plot (on a log-log scale) $\Delta J_{light}$ at $V=0$ as a function of $\delta$ (which is proportional to the illumination intensity). The linear relation is evident, and is different than the power-law dependence on intensity expected from the thermo-electric effect. 

% to add after review - discuss natural trade-off between pushing the molecular resonance further away from Fermi energy and the signal - compare Fig. 2 with its inset (factor 100 or so for a shift from 0.77eV to 1.4 ev>

\section{Summary}
In conclusion, we have shown that our analytical prediction of the electron distribution under continuous wave illumination can be used to interpret recent experiments (RW20). This sheds new light on these experiments, and demonstrates that they seem to measure directly the NTC contribution to the distribution function, however, surprisingly, not as they originally interpreted their data. We suggest further experiments that can be analyzed within our theory, and the necessary improvements to state-of-the-art theories for the electron non-equilibrium distribution and current through a MJ, thus, providing a direct route to solving one of the outstanding questions in plasmonic systems, namely, the form of the electron distribution under continuous illumination. {\XYZ Finally, the methodology presented here can be used for as a starting point for a theoretical treatment of other experiments where nano-scale transport is coupled to plasmonic effects~\cite{evans2019remote,Bouhelier_PL_2019, Cui2020Electrically,Cui2021Thousand,Bouhelier_PL_2019,Bouhelier_PL_2020}. }

% unfortunately, the following is not directly relevant anymore - The observation associated with the spatial averaging raises a question mark about the validity of the claims in~\onlinecite{Khurgin-Faraday-hot-es} that the generated non-thermal electrons are concentrated in a sub-nm region near the metal interface. 

% how they heat their sample in the control experiment?

% Things that could be said somewhere:
% \begin{itemize}
%     \item they used high quality of metal film. roughness not expected to raise the number of hot e's too much

%   \item should we include the eDOS in a proper calculation to see if it helps?
 
%   \item shoulders in Figs. 1-3 look too flat. how come?

% \end{itemize}

% \begin{acknowledgement}
% The authors thank G. Bartal, H. Reddy, P. Reddy and K. Wang for fruitful and illuminating discussions. %YD and YS are not funded by any oil and gas companies, because we support the climate!
% \end{acknowledgement}

 \begin{suppinfo}
Supplementary Information includes additional transport calculation details, fit parameters and thermal calculations. This material is available free of charge via the internet at http://pubs.acs.org.

\end{suppinfo}

%\bibliography{my_bib.bib}

\begin{figure}[h]
\centering{\includegraphics[width=0.8\textwidth]{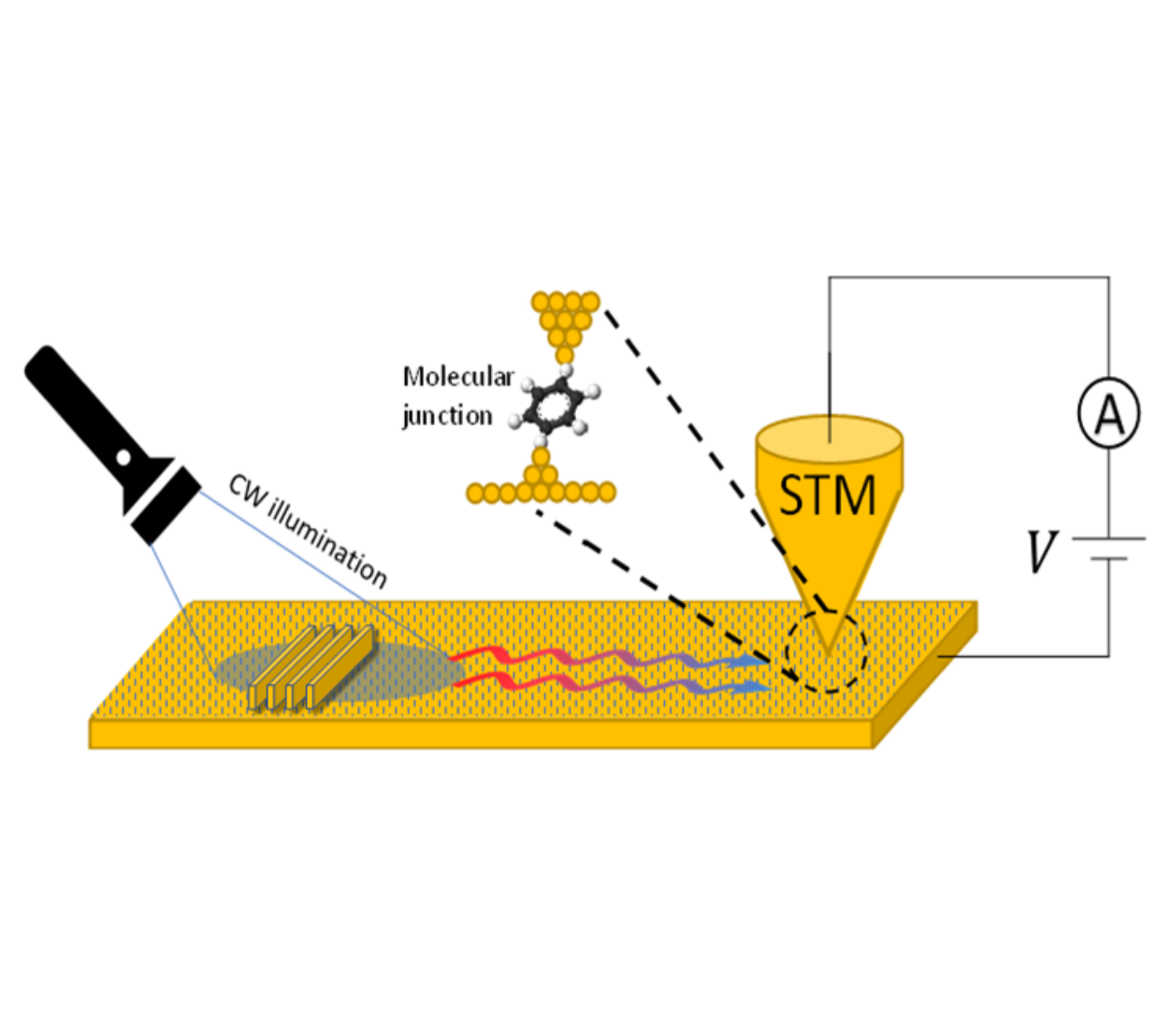}}
\caption{TOC grasphics}
 \label{fig:TOC}
\end{figure}

\providecommand{\latin}[1]{#1}
\makeatletter
\providecommand{\doi}
  {\begingroup\let\do\@makeother\dospecials
  \catcode`\{=1 \catcode`\}=2 \doi@aux}
\providecommand{\doi@aux}[1]{\endgroup\texttt{#1}}
\makeatother
\providecommand*\mcitethebibliography{\thebibliography}
\csname @ifundefined\endcsname{endmcitethebibliography}
  {\let\endmcitethebibliography\endthebibliography}{}

\end{document}

\section{Molecular parameters extraction from experimental data in RW20}
In their original manuscript RW20, the analysis and extraction of ``hot''-carrier contribution to the current (and essentially all other analysis) is based on the Landauer formula. Here, we demonstrate that the experimental data of RW20 is, in fact, inconsistent with their choice of molecular parameters. 

Within the Landauer formalism, the relation between current, voltage and transmission is given by (Eq. S16 in RW20)
\begin{equation} \label{Landauer}
J(V) = \frac{2e}{h} \int d\E \mathcal{T}(\E) \left(f^T(\E) - f^T(\E + e V)\right)~~,
\end{equation}
where $f^T$ is the equilibrium Fermi distribution, $V$ the bias voltage, $e$ the electron charge, $h$ Planck's constant. $\mathcal{T}(\E)$ is the transmission function, which is taken to be of Lorentzian form $\mathcal{T}(\E) = \frac{\Gamma^2}{\Gamma^2 + (\E - \E_0)^2}$, where $\E_0$ is the energy of the frontier molecular orbital and $\Gamma$ is the level broadening (see Eq.~S18 in RW20). 

In RW20, the authors used a numerical differentiation methodology (suggested in, e.g.,~\onlinecite{capozzi2016mapping}) to extract the molecular parameters $\Gamma = 2.6$ meV and $\E_0 = 0.18$ eV for the L1 molecule. These parameters were used for the NTC analysis, and for showing, for instance, that the thermoelectric response should be small. 

\begin{figure}[t]
\centering{\includegraphics[width=0.8\textwidth]{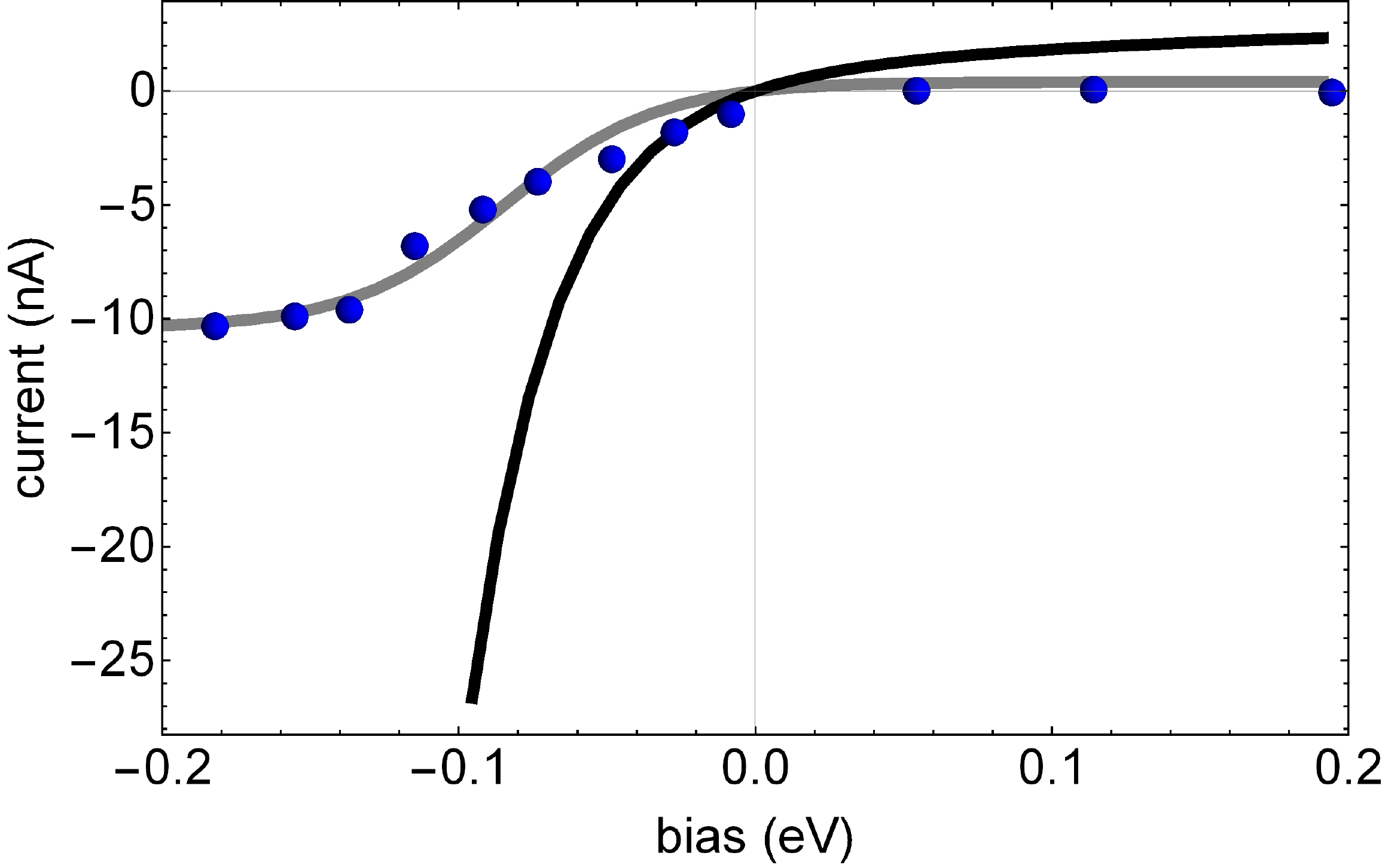}}
\caption{(Color online) Experimental I-V curve (blue circles), extracted from RW20, along with the theoretical I-V curve from the molecular parameters of RW20 (black line), showing the discrepancy between the suggested molecular parameters and the experimental results. Gray and yellow lines are fits to different molecular parameters (see text).}
\label{S-fig1}\end{figure}

In Fig.~\ref{S-fig1} we plot the current, extracted from Fig.~S15A of RW20 (blue circles). On top of it, the black solid line is $J(V)$ as obtained by using Eq.~\ref{Landauer} described above with the numerical parameters found in RW20. It is rather clear that these parameters do not lead to a good fit between the theoretical description and the experimental data. In fact, a much better fit can be found (solid gray line), using the parameters $\Gamma = 0.0466$ meV and $\E_0 = 0.087$ eV. However, these parameters seem to be quite unrealistic (or at least quite exceptional), because of the extreme smallness of $\Gamma$. For comparison, in Fig.~\ref{S-fig1} we also plot the current for $\Gamma = 0.04$ eV and $\E_0 = 0.28$ eV (yellow line). The point here is that the molecular parameters are largely unknown and can be fitted in various ways.  

\begin{figure}[t]
\centering{\includegraphics[width=0.8\textwidth]{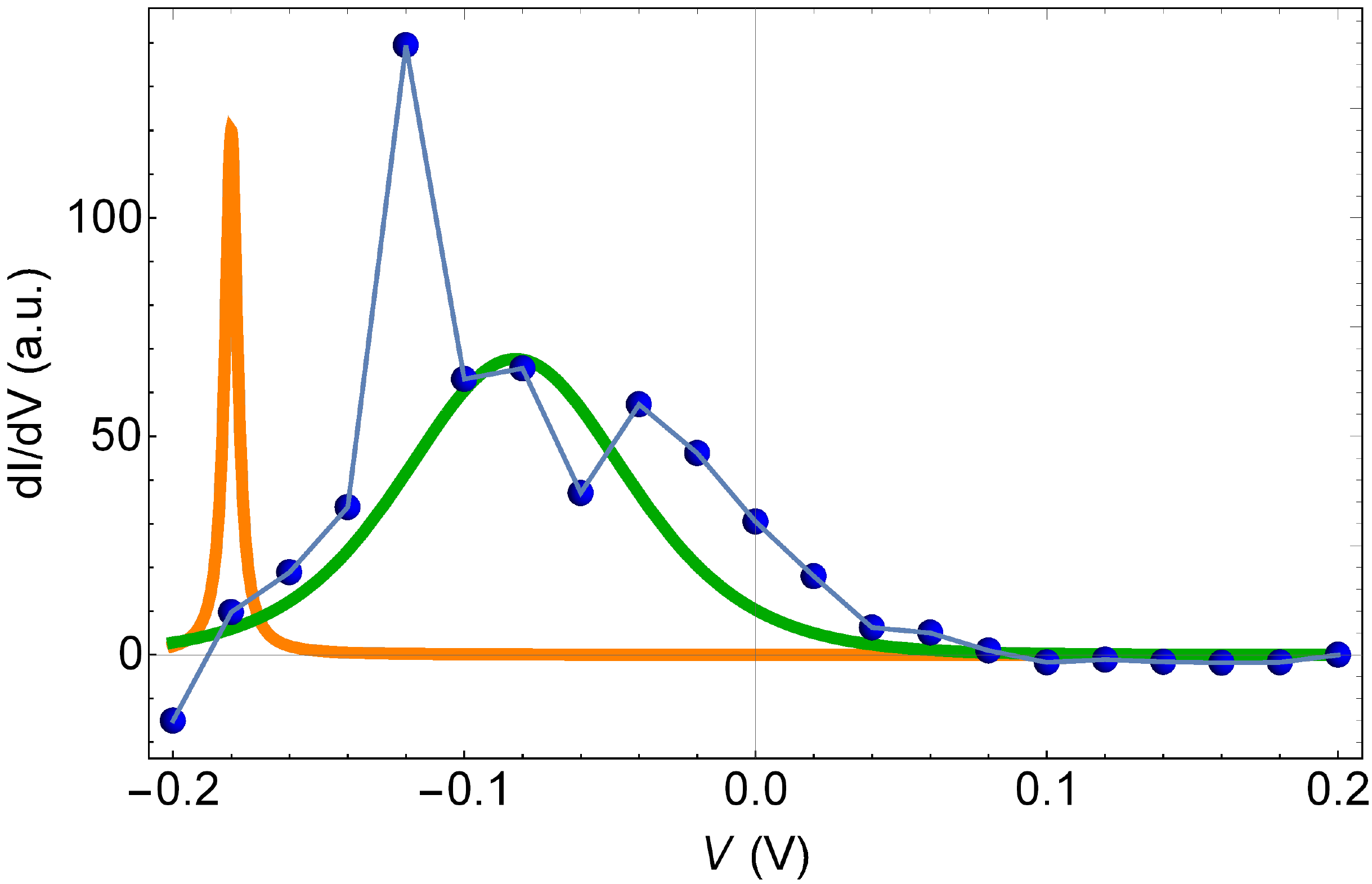}}
\caption{(Color online) Experimental transmission curve (I-V derivative, blue circles), extracted from RW20, along with the theoretical transmission curve from the molecular parameters of RW20 (orange line), showing the discrepancy between the suggested molecular parameters and the experimental results. Green line is a fit to different molecular parameters (see text). } 
\label{S-fig2}\end{figure}

The discrepancy between the molecular parameters and experiment in RW20 can also be seen by looking at the numerical derivative of the current, which can be related directly to the transmission function (Eq.~S17 in RW20). In Fig.~\ref{S-fig2}, we plot the numerical derivative of the I-V curve (blue circles). The orange line is the Lorentzian function with the molecular parameters suggested in RW20. The green line is the full derivative of the Landauer expression $J(V)$ using the suggested (non-realistic) parameters above. 

%\section{Fit to $\Delta J_{light}$ for {\bf L2} molecule}
%As described in the main text inset to Fig.~2), the experimental $\Delta J_{light}$ can be fitted reasonably well if the molecular parameters are taken from other measurements (Ref.~26 in the main text) and are not used as fit parameters. From Sec.~1 above, it seems that there may be a large variability of the molecular parameters between experiments. With this in mind, we performed a fit to the experimental $\Delta J_{light}$ for {\bf L2} molecule, using both $\Gamma$ and $\E_0$ as fitting parameters. We find that an excellent fit can be achieved for $\Gamma = 0.014$eV, $E_{LUMO} = 0.377$eV and $\delta_E^{(0)} = 5 \times 10^{-5}$eV$^2$ (see Fig.~\ref{S-fig3}). These molecular parameters are different from what was previously measured for the same molecule (Ref.~26 in the main text), but are quite reasonable in general, and cannot be ruled out. 

%\begin{figure}[t]
%\centering{\includegraphics[width=0.8\textwidth]{S-fig3.pdf}}
%\caption{(Color online) Fit between the experimental $\Delta J_{light}$ for the {\bf L2} molecule (blue circles, extracted from RW20), and our theory (Eq.~(3) {\bf still??} in the main text) with molecular parameters $\Gamma = 0.014$eV, $E_{LUMO} = 0.377$eV and $\delta_E^{(0)} = 5 \times 10^{-5}$eV$^2$ (black squares), showing excellent agreement between theory and experiment. } 
%\label{S-fig3}\end{figure}

\section{Fit for $\Delta J_{light}$ at different surface temperatures} \label{sec:dJ_fit_surface_T}
As was shown in the main text (Fig. 2(c)-(d)), an excellent fit between experimental values of $\Delta J_{light}$ can be obtained by allowing the molecular parameters and the surface temperature to serve as fit parameters. While the former can be (at least conceptually) determined in control experiments, unfortunately, the latter parameter is hard to determine independently. Indeed, from the experimental point of view, the surface temperature below the STM tip is extremely hard to measure (and indeed, was not measured directly in RW20). From the numerical point of view (see SI Section 3 for details), the uncertainty in the various system parameters (namely, the size and shape of STM tip, its distance from the slab, the molecule size etc.) gives rise to a large uncertainty in the resulting temperature. 

\begin{figure}[t]
\centering{\includegraphics[width=0.8\textwidth]{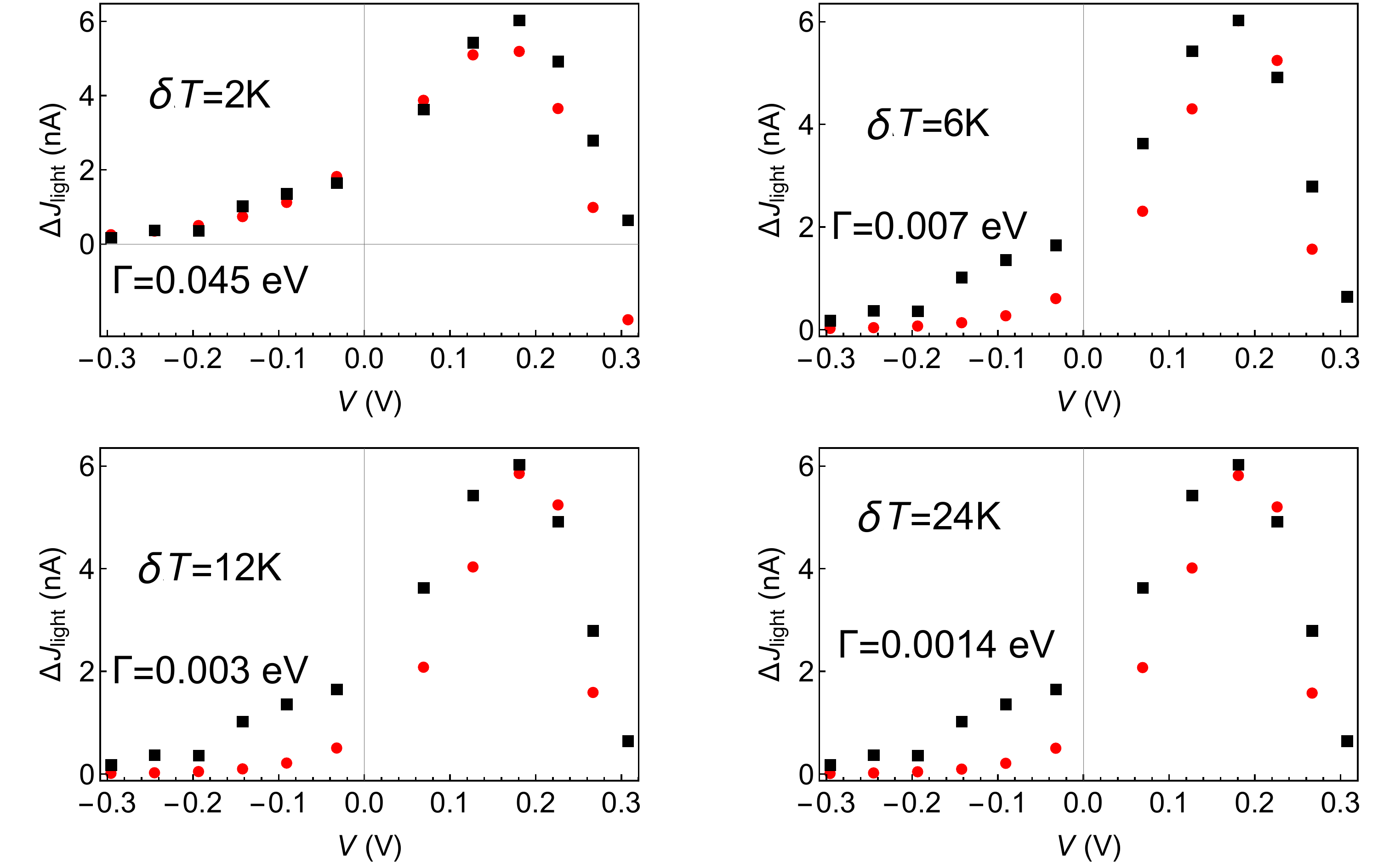}}
\caption{(Color online) $\Delta J_{light}$ vs voltage. Data (black squares) extracted form RW20. Fits (red circles) are obtained by setting the temperature and molecular energy constant, and varying only the level broadening $\Gamma$. } 
\label{S-fig-highT}\end{figure}

However, this uncertainty does not hamper the relevance of our analysis. Indeed, one can aim at fitting the experimental values of $\Delta J_{light}$ by fixing the temperature and changing the molecular parameters% (which are also essentially unknown) - commented, because they could be perfectly well know...
. In Fig.~\ref{S-fig-highT} we show fits to the $\Delta J_{light}$ data from RW20 (solid squares) and fits using slab temperatures (which we define as $T_{slab} = T_{dark} + \delta T$) of $\delta T = 2$ (as in the obtained fit in Fig.~1(c)-(d)) but also of $\delta T = 6,12,24$ K, the latter corresponding to the value obtained in our numerical simulations, see Fig.~\ref{fig:dT_6nm}(c). The theoretical fits (red circles) have a fixed molecular energy $\E_0 = 0.15$ eV, and {\sl only} allow for the level broadening $\Gamma$ to be changed. Simply put, these are fits with only a single free parameter (and still the fits are quite reasonable). % The values of the level broadening are marked in the figure. 

\section{Numerical simulations}
Understanding the optically-induced current in the MJs requires knowing the local values and spatial distribution of the electric field and temperatures in the tip-slab structure. To this end, we simulated a Au tip ($800$nm in diameter, a $800$nm taper and $1$nm tip curvature) above a thin Au slab ($6$nm and $13$nm thick, as in RW20); the tip-slab separation was chosen to be $1$nm. The overall computational domain size was cylindrical, with $1 \mu$m height and $1.6 \mu$m diameter; it was surrounded by Perfectly Matched Layers (PMLs). For all the simulations we verified that the results are not sensitive to the size of the simulation domain nor to the mesh size.
% tip at the center to make sure the PML works best.

% see~\onlinecite{Shalaev_tapered_PRL,Antonio_kissing_spheres_PRL,Ciraci:2012_Science,angela-NPoM-QNM} for similar structures). 

\subsection{Electromagnetic simulations}
Optical properties of Au were taken from Ref.~\citenum{optic_constant_noble_metals}, and the refractive index of the fused silica and matching oil was set to $1.45$; as in RW20, we do not account for the molecular layer index on top of the Au slab. Since the SPP generation by the grating was described in RW20 in great detail, here we refrain from modelling it. Instead, we just adopt the wavelength ($830$nm), intensity and profile of the (excited) long-range SPP as detailed in RW20 to simulate the SPP interaction with the slab (not studied in RW20); in particular, since it has been shown in RW20 that approximately 15\% of the incident beam energy was coupled to the SPP by the grating, we set the input SPP field in the simulation to be $\sqrt{0.15} E_\textrm{inc}$, where $|E_{inc}| = 3.73 \times 10^5$ V/m is the incident field amplitude in the experiment\footnote{This amplitude is obtained from to the incident power using
$$ 
\dfrac{P_0}{A} = \dfrac{1}{2}\varepsilon_0 c_0 \sqrt{\varepsilon_{\textrm{SiO}_2}} |E_0|^2 \approx 286 MW/m^2, 
$$
where $P_0 = 6.6$ mW is the power of the incident beam in RW20, $A$ is the spot area (for a 5.6 $\mu$m beam diameter), $\varepsilon_0$ is the vacuum permittivity, $c_0$ is the speed of light in vacuum and $\varepsilon_{\textrm{SiO}_2}$ is the relative permittivity of the fused silica. Note that the value reported in RW20 for the intensity ($0.3 GW/m^2$) is slightly different but not not exactly the same.}; %and hence, set the input field to the simulation to be $\sqrt{0.15} E_{inc}$; 
we ignore the residual direct illumination of the tip by the wide input beam.
% simulation shows ?? the generated SPP travelling upwards along the tip and the short range SPP scattered from the tip onto the slab; We verified that the latter's propagation constant, spatial extent and symmetry are produced by the simulation correctly 

\begin{figure}[h]
\centering
\includegraphics[width=0.9\textwidth]{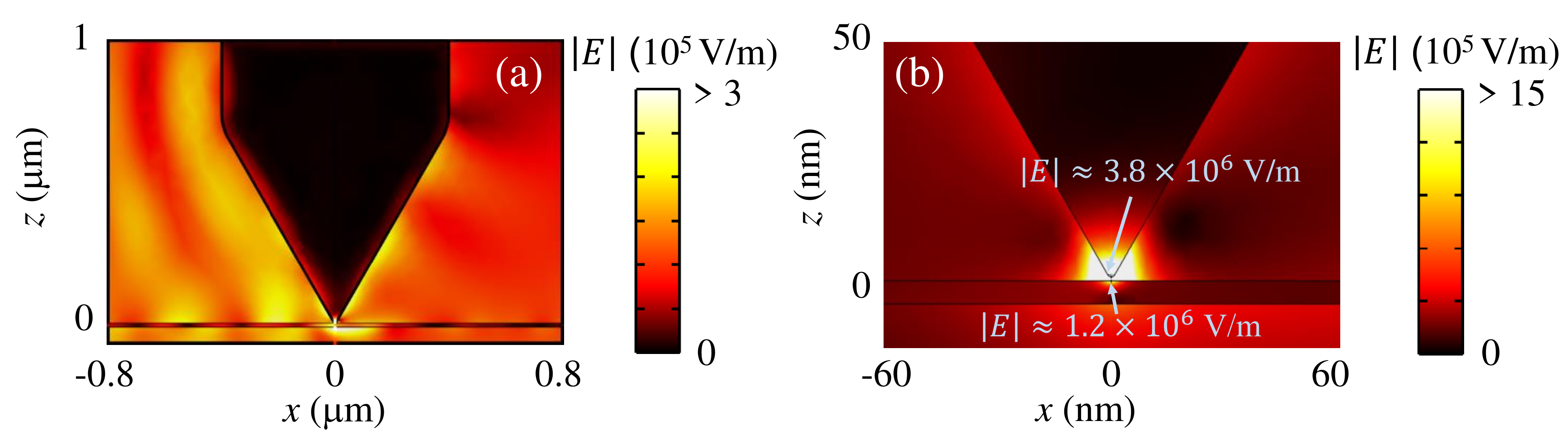}
\caption{(Color online) (a) Simulation results of the local field amplitude $|E|$ for the SPP-excitation experiment. (b) shows the the zoom-in of the region of the metal tip. } 
\label{fig:normE_6nm_spp}
\end{figure}

In Fig.~\ref{fig:normE_6nm_spp} we plot the amplitude of the local field $|E|$. One can see that the amplitude of the local field at the tip edge reaches $|E| \sim 3.8 \times 10^6$ V/m, about $3$-fold higher than in the slab ($|E| \sim 1.2 \times 10^5$ V/m) right under it. The enhancement with respect to the incident field is a result of the vertical and lateral focusing induced by the presence of the tip and the presence of a vertically-oriented electric field component in the SPP-based illumination in RW20~\cite{reddy2020determining}. This enables the excitation of the strongly-localized (dipolar) modes of the nano-cavity created by the tip and slab. Since, in addition, the STM tip is sharp, the local field in it is higher than in the slab.

For this value of local field, and assuming~\cite{BTl} $K = 8 \times 10^{12} [\textrm{eV}^{-2} \cdot \textrm{s}^{-1}]$ and using~\footnote{We set the Fermi energy to $\e_F = 5.1$eV, $\epsilon_{Au}(\lambda = 830nm) = -28.8 + 1.6i$, $\epsilon_{host} = 1.5^2$, and the electron density to $n_e = 5.9 \times10^{28}$ m$^{-3}$. The photon frequency is $\hbar \omega_L = 1240 / 830$ eV. } $R = \dfrac{4 \epsilon_0 \epsilon''_m}{3\hbar} \cdot \dfrac{\e_F}{\hbar \omega_L n_e} = 1.03 \times 10^{-5}$ [m$^2\cdot$V$^{-2}\cdot$s$^{-1}$], we get that $\delta_{E,top}(1) = 1.15 \times 10^{-4}$ and $\delta_{E,bottom}(1) = 1.1 \times 10^{-5}$ . This value is an order of magnitude lower than the value obtained from the fit. It should, nevertheless, be noted that the value of $K$ has certain uncertainty (e.g., compare to~\onlinecite{delFatti_nonequilib_2000}), that the exact geometry of the tip is not known exactly, and that the residual direct illumination of the tip by the wide input beam will affect the tip and slab field (and temperatures) as in the bottom illumination control, i.e., %(ignoring interferences; who knows if this is justified...) 
it would modify the reported values and reduce the difference between the local fields. Furthermore, it is critical to recall that while the theories for the electron non-equilibrium distribution and for transport through a MJ both assume that the fields are uniform, these quantities are far from being so in the specific configuration we study, thus, putting another hard-to-quantify ``error bar'' on the theoretical values. In fact, one may wonder whether it would be appropriate to spatially average the electric field, temperature and electron distribution in order to obtain a more suitable theoretical prediction; the exact way of how to do this is a challenge for a future study. Either way, all this means that the difference we find between the value of the fit of the experimental data and the simulations is quite reasonable.

%One can see that the amplitude of the local field at the tip edge reaches $|E| \sim 3.8 \times 10^6$ V/m (the amplitude enhancement reaches $|E|/|E_0| \sim 10.2$), about $3$ fold higher than in the slab ($|E| \sim 1.2 \times 10^5$ V/m and $|E|/|E_0| \sim 3.2$) right under it. This enhancement is a result of the strong vertical and lateral focusing induced by the presence of the tip. {\bf explain the delta computation?}

\begin{figure}
    \centering
    \includegraphics[width=0.9\textwidth]{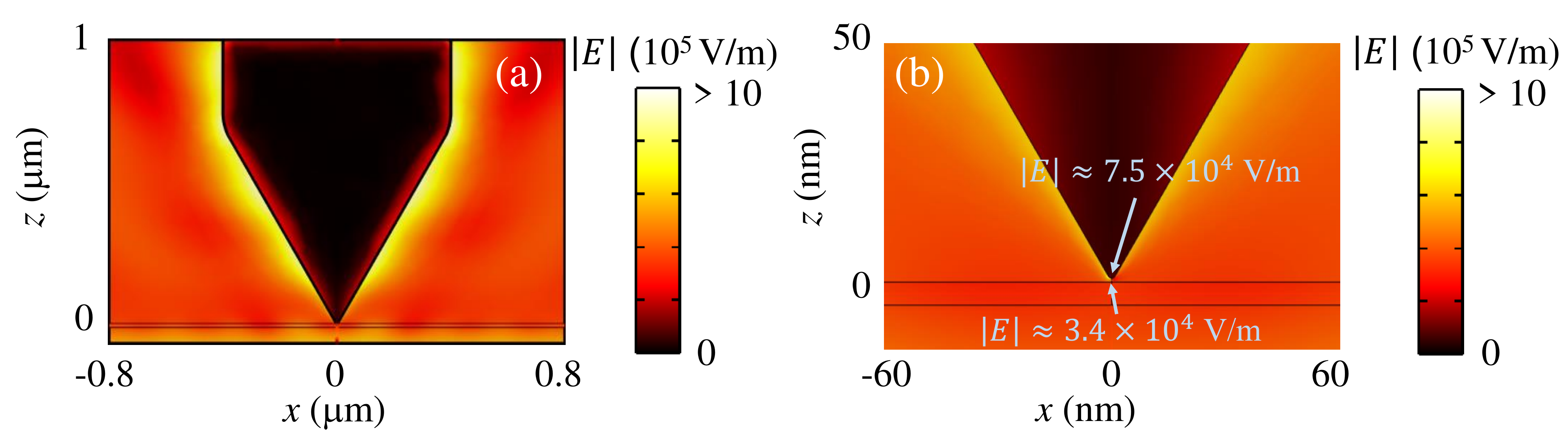}
    \caption{(Color online) (a) Simulation results of the local field amplitudes $|E|$ for the bottom illumination experiment. (b) shows the the zoom-in of the region of the metal tip.}
    \label{fig:normE_6nm_norminc}
\end{figure}

The control experiment is described in Fig.~\ref{fig:normE_6nm_norminc}. %Here, $|E_0|$ is again the amplitude of the incident electric field. 
%For simplicity, we simulated a plane wave ($|E_0| = 3.94 \times 10^5$ V/m) incident from the silica substrate upwards, since the $5.6 \mu$m Gaussian beam used in RW20 is much wider than the tip. One can see that in this case, the input field is mostly transmitted through the thin slab and that the fields excited in the tip are of a depolarizing nature. This makes the field across the tip ($|E| \sim 7.50 \times 10^4$ V/m $|E|/|E_0| \sim 0.19$) significantly lower than the field in the slab ($|E| \sim 3.38 \times 10^4$ V/m and $|E|/|E_0| \sim 0.86$). {\bf put explanation here (instead of in text)?}
For simplicity, we simulated a plane wave incident from the silica substrate upwards, since the $5.6 \mu$m Gaussian beam used in RW20 is much wider than the tip. One can see that in this case, the input field is mostly transmitted through the thin slab (as expected) and that the fields excited in the tip are of a depolarizing nature. This makes the field across the tip ($|E| \sim 7.5 \times 10^4$ V/m) significantly lower than the field in the slab ($|E| \sim 3.38 \times 10^4$ V/m). The reason for this different behaviour is that for this control experiment, the vertical field component is absent from the illumination, so that there is no coupling to the dipolar modes and only the minor polarization is excited.

\subsection{Thermal simulations}
For these simulations, the thermal conductivity of the index matching oil (propylene carbonate) is set to $0.16$ W/m$\cdot$K~\cite{kappa_propylene_carbonate}, $310$ W/m$\cdot$K for Au, and $1.38$ W/m$\cdot$K for the fused silica substrate, as in RW20. In addition, we account for the finite interface (Kapitza) conductivity; its values on the metal/oil interface and on the metal/fused silica interface are set to be $200$ MW/(m\textsuperscript{2}$\cdot$K) and 2 GW/(m\textsuperscript{2}$\cdot$K), respectively~\cite{Del_Fatti_kapitza_glass_Au,Stoll_environment,AuPd_water_interface_thermal,Au_nanorod_solvent_interface_thermal}. This can be done by assigning a thin layer feature to the metal/oil interface (and the metal/silica interface) in the simulation.

The heat source is separated into two parts. For the metal tip and the metal slab within the simulation domain, the heat source is computed from the absorbed energy density using the EM field obtained from the EM simulation described above and using Poynting's Theorem. For the metal slab in the PML, the heat source is the absorbed energy density calculated using the incident field only; it is properly transformed using transformation optics (see~\onlinecite{Un-Sivan-size-thermal-effect}) such that the PML region (a size of $250$nm in the simulation) models a size of $30 \mu$m in radius on the sample. Since we do not simulate the grating, we follow RW20 and set the heat source to have a Gaussian profile centered at the position 3.8 $\mu$m away from the tip; the magnitude of the heat source was chosen to be 8 times larger the absorbed energy density in the gold film under normal incidence using the 5.6 $\mu$m Gaussian beam such that the maximal heating reaches $67$K, as reported in RW20 (Fig. S23; these simulations seem to have been performed in the absence of the grating and the metal tip). This value is %8 times higher than the absorption of the incident illumination only, 
made possible by the resonant nature of the grating. % {\bf but if they got the boundary condition incorrectly, then, this is exaggerated... should we simulate it? should we ask it of them?} 

% also describe your 2D tests...??

% Temperature distribution hints on charge distribution - spatial averaging washes out effect on tip (although local field at tip is stronger).

\begin{figure}[h]
\centering
\includegraphics[width=0.7\textwidth]{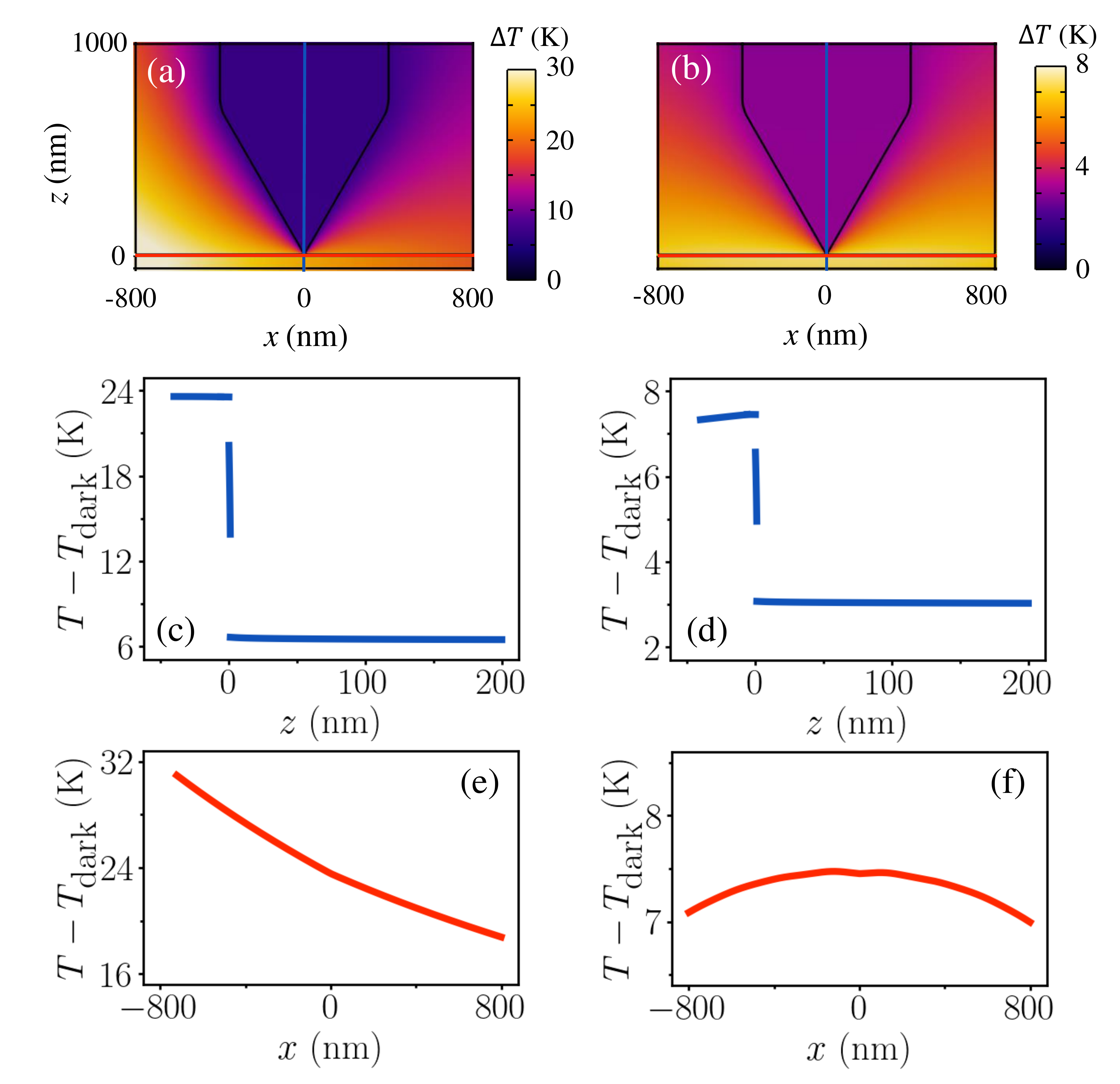}
\caption{(Color online) Simulation results of the temperature profile for the experiments with (a) SPP-excitation and (b) bottom illumination. (c)-(d) The corresponding temperature rise profile along the $z$-axis (labeled by blue solid lines in (a)-(b)). The temperature jumps shown in (c) and (d) are due to the the finite interface (Kapitza) conductivity. (e)-(f) The corresponding temperature rise profile along the $x$-axis (labeled by red solid lines in (a)-(c)); the temperature increase leftwards in (e) is due to the presence of the strong absorption at the grating.} \label{fig:dT_6nm}
\end{figure}

In Fig.~\ref{fig:dT_6nm} we plot the temperature profile in two cases (SPP-excitation and bottom illumination). Although the local field at the tip was shown above to be higher, we observe that for the SPP-based illumination, the tip temperature reaches $\sim 6$K, about $\sim 18$K colder compared to the slab region right under the it. This somewhat surprising finding is explained by the smallness of the strongly illuminated tip volume, which simply does not generate as much heat as the somewhat weaker but beam-wide heat source in the slab/grating. Instead, the tip heating is mediated primarily by the liquid, which is itself heated by the heated grating. For that reason, the Kapitza resistance has a negligible effect on the temperature distribution.

% transverse gradients - should not be severe were 40% in Baffou's rod; but here the illumination is uniform, s the effect would be far weaker.

Most importantly, we note that the tip dimensions in RW20 were much larger than those we were able to simulate. To account for that, we verified that the temperature rise in both slab and tip is higher for thinner tips, significantly more so in the tip. This finding supports the assumption in RW20 that the tip was not heated, and that the slab under it was heated by only a few degrees. In fact, if we constraint the fits by setting the temperature of the bottom electrode (at different temperatures, even up to 20K, taking into account that simulation details and unknown geometrical parameters such as tip size etc may change the numerical estimate of the temperature), we can still obtain excellent fits to the data (see SI Section~\ref{sec:dJ_fit_surface_T}), albeit with different molecular parameters (mostly the level broadening $\Gamma$).

For the bottom illumination (control experiment), the temperature rise in the slab and the tip are $\sim 7.8$K and $\sim 6$K, respectively. This explains Fig.~S12A in RW20, which showed no change in current under bottom illumination. The reason is that such a small increase in temperature does not affect the current (as is evident from Fig.~S3(a) above).  

%tip is $\sim 6$K, about $1.8$K lower than in the slab segment right under it. {\bf so what is the significance of this? connect to sentence below... that the heating is comparatively weaker in the control experiment compared with the SPP-based excitation experiment... a bit problematic (because of the factor 8)... % discuss temperature uniformity in slab?

%{\bf edit the following once we get results with Kapitza - } 
%Accounting for the residual illumination of the tip by the wide input beam is likely to have made this difference ?? , so that our simulations seems to be in ?? with the value obtained from the fit... 

%Yoni - you said that for the control the 1.8K we got gives 2 nA (via (S1)??), whereas they measured $1 \pm 1$nA. is that right?}

% For the resistive-heating controls experiments, we observe that ... {\bf Yoni - please discuss significance of these results}

% We account for heat transfer mediated by evanescent fields. Yoni - you said this effcet is weaker by a factor 1000 compared with the molecule, which is itself negligible?

\subsection{Thermal control experiments in RW20}
\begin{figure}[t]
\centering{\includegraphics[width=1\textwidth]{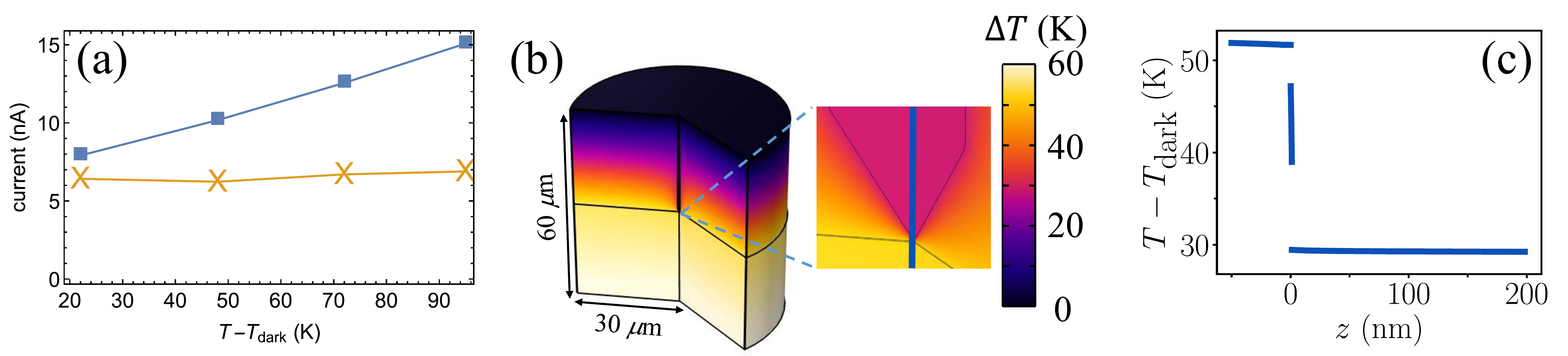}}
\caption{(Color online) (a) Experimental (crosses) and theoretical (squares) current as a function of temperature (above room temperature) at constant bias of $0.1$V. Data extracted from RW20 Fig.~S12. (b) The simulation results of the temperature profile for the thermal control experiment. In the simulation, we set a uniform heat inflow boundary condition to the bottom surface (30 $\mu$m away from the metal slab). The amount of in-flowing heat was adjusted so that the metal slab reached 72$^\circ$C. In addition, we set a zero-heat-flux boundary condition to the side boundary (30 $\mu$m away from the tip); this minimizes the horizontal temperature gradients, as expected due to the very large size of the heat source at the bottom of the fused silica substrate. Finally, we set the temperature of the top surface (30 $\mu$m away from the metal slab) to be $T_\textrm{dark}$; no sensitivity to this value was found. (c) The corresponding temperature rise profile along the $z$-axis (labeled by the blue solid line in (b)).} 
\label{S-fi_IT1}
\end{figure}

In Fig.~S12 of RW20, the authors present a measurement of the current at a bias voltage of 0.1V, for 4 different temperatures above room temperature ($\Delta T$). These measurements are performed by resistive-heating of the Au slab from the bottom. Their data (extracted from Fig. S12 of RW20) is shown in Fig.~\ref{S-fi_IT1} (crosses), and shows essentially no temperature dependence. 

However, this result seems to be inconsistent with the molecular parameters provided in RW20. To see this, we evaluate the current at 0.1V bias voltage using the Landauer equation~(\ref{Landauer}), using the parameters provided in RW20 (see SI Section 1). The resulting current is plotted in Fig.~\ref{S-fi_IT1} (filled squares), and shows a clear rise of current with temperature, of about $\sim 10$ nA. 

One possible origin for this discrepancy is that the top electrode (the STM tip) is much colder than the slab (so that the actual temperature difference to use in Eq.~(\ref{Landauer}) is much lower). However, in order for the measured data to match the theoretical calculation~(\ref{Landauer}), we find that the top electrode must remain at room temperature even when the bottom electrode is heated up by $90$K. This seems to contradict our numerical simulations, which show that a rise in the temperature of the bottom slab is accompanied by heating of the tip to almost the same level (Fig.~\ref{S-fi_IT1}(b)-(c)). 

Another possibility is that the molecular parameters provided in RW20 are somewhat inaccurate. As seen in SI Section 1 above, this is not an unlikely possibility. Indeed, for $\E_0 \sim 0.4$eV and $\Gamma \sim 0.01$eV the theoretical I-V curves match the experimental ones (but only at the negative bias regime!), and give very small change in current ($<1 $nA rather than $\sim 5 nA$) over the temperature range measure in RW20 Fig.~S12. However, such molecular parameters are inconsistent with previous calculations~\cite{wang2019charge}.  

% so bottom line - the wrong molecular parameters may not be the only problem...

\subsection{Conclusions from the simulations}

The trends described above are rather insensitive to the exact geometry of the tip or to the slab thickness, and will occur also for somewhat different configurations (e.g.~\onlinecite{angela-near-far-field,Antonio_Nat_Comm_2020}), as long as the tip is not completely flat. Thus, several conclusions can be drawn

\begin{enumerate}
    \item The SPP-based excitation employed in RW20 creates the unique conditions enabling the challenging need to create different electric fields and temperatures across the MJ.
    
    \item In particular, they create a sufficiently higher temperature on the slab compared with the tip, and potentially even negligible heating of the tip. This supports the assumptions on which the analysis of the L1 molecule in RW20 were based.
    
    \item In addition, the simulations show that the electric fields in the MJ exhibit an opposite trend with respect to the temperature, namely, a stronger field at the tip compared with the slab right under it. This enabled us to demonstrate a qualitative and quantitative match between fit to the experimental data and the calculations for the L2 molecule, providing what seems to be the first ever observation of the number high energy non-thermal electrons under steady-state illumination. 
    
    \item This agreement is obtained under the assumption that the difference between the electron and phonon temperatures is negligible, in agreement with theoretical predictions~\cite{Abajo_nano-oven,Dubi-Sivan,Dubi-Sivan-Faraday} based on conventional values for the $e-ph$ coupling coefficient~\cite{delFatti_nonequilib_2000,Brown_PRB_2016}. In that respect, there is no support for the claim that the steady-state $e-ph$ coupling coefficient should very different from its value extracted from ultrafast experiments (see~\cite{Sheldon-JPCC-2020}).
    
    \item On the other hand, our simulations also show that the bottom illumination (i.e., control) experiment, does not create the same conditions as those in the SPP excitation as far as the electric field (hence, the true non-thermal electron distribution) is concerned.
\end{enumerate}

Beyond all these conclusions, it is clear that the state-of-the-art theory for the electron non-equilibrium in metals has to be extended to enable it to describe correctly changes to the distribution due to interband transitions (not necessary for the current work), as well as inhomogeneities in the electric field. Similarly, the Landauer formula also should be extended to account for non-uniform temperature (and even electron distribution), to enable it to describe the specific experimental situation at hand. These are challenges for future development.

%\bibliography{my_bib.bib}
\providecommand{\latin}[1]{#1}
\makeatletter
\providecommand{\doi}
  {\begingroup\let\do\@makeother\dospecials
  \catcode`\{=1 \catcode`\}=2 \doi@aux}
\providecommand{\doi@aux}[1]{\endgroup\texttt{#1}}
\makeatother
\providecommand*\mcitethebibliography{\thebibliography}
\csname @ifundefined\endcsname{endmcitethebibliography}
  {\let\endmcitethebibliography\endthebibliography}{}